\documentstyle[12pt]{article}
\def\presentation{
\oddsidemargin 0in
\evensidemargin 0in
\marginparwidth .75in
\marginparsep 7pt
\topmargin 0in
\headheight 12pt
\headsep .25in
\footheight 18pt
\footskip .35in
\textheight 9in
\textwidth 6in
\columnsep 10pt
\columnseprule 0pt
}
\presentation

\title{General Solution of the Consistency Equation}
\author{Michel Dubois-Violette\thanks{L.P.T.H.E. Universit\'e Paris XI
 B\^atiment 211 F-91405 ORSAY CEDEX}
\and Marc Henneaux\thanks{
Ma\^\i tre de Recherches au Fonds National de la Recherche Scientifique
\newline
Facult\'e des Sciences / U.L.B. Campus Plaine C.P. 231 B-1050
BRUXELLES}
 \and  Michel Talon$^{\small{\ddag}}$
\and Claude-Michel Viallet\thanks{L.P.T.H.E. Universit\'e Paris VI
 Bo\^{\i}te 126 /4 place Jussieu/ 75252 PARIS CEDEX 05}}
\date{\today}
\def\a{{\cal A}}
\def\b{{\cal B}}
\def\h{{\cal H}}
\def\d{\delta}
\def\g{{\cal G}}
\def\k{{\cal K}}

\begin{document}
\bibliographystyle{perso}

\begin{titlepage}
\renewcommand{\thepage}{}
\maketitle
\vskip 2cm
\begin{abstract}
We produce  the general solution of the Wess-Zumino consistency
condition for  gauge theories of the Yang-mills type, for any ghost
number and form degree. We resolve the problem of the cohomological
independence of these solutions. In other words we fully
describe the local version of the cohomology of the BRS operator,
modulo the differential on space--time. This in particular includes
the presence of external fields and
non--trivial topologies of space--time.
\end{abstract}
\vfill
PAR-LPTHE 92/19 \hfill\newline
LPTHE-ORSAY 92/33 \hfill
Work supported by CNRS
\end{titlepage}
\renewcommand{\thepage}{\arabic{page}}

\section{Introduction}
We investigate the Wess-Zumino consistency equation~\cite{WZ71} in non abelian
gauge theories on an arbitrary space-time. We produce their general solution,
taking due account  of the possible non-trivial cohomology of space-time.

Motivated by the problem of quantum anomalies in renormalization
theory~\cite{BeRoSt76,Fad84}, and more recently by the discovery of
their central r\^ole in topological quantum field theories~\cite{Wit88}
there has been a number of publications on the
subject~\cite{Di76,Di77,St76,Di79,St83,Bau83,ThM84,dvtv85,dvtv85b,%
dvtv86,Ban86,Ban87,Di90,BrDrKr90,BrDrKr90b,He91,hdtv91,Di91},
many of these being a step towards the solution.
We obtain the full resolution by an elaboration of our results
of~\cite{dvtv85b} and~\cite{hdtv91}.

The proper way to look at the  Wess-Zumino consistency equation is to
view them as a problem of {\em local} cohomology~\cite{BeRoSt76,St76,BoCo83}.
In the general problem, one has to accept the presence of external
fields,
and the existence of a non trivial cohomology for space--time.

The specificity of the cohomological problem we want to solve is the
following: we deal with the algebra $\a$ of form--valued
polynomials in the fields and a finite number of their derivatives.
Two differentials are defined on $\a$, the exterior derivative $d$ on
space--time, and the
B.R.S. operator $\d$. These two differentials verify:
\begin{equation}
\d d + d \d = \d^2 = d^2 =0 \label{ddel}
\end{equation}
The consistency condition on $Q \in \a$ means there exists $R\in\a$ such that
\begin{equation}
\d Q + d R =0
\label{consist}
\end{equation}
and any solution of (\ref{consist}) of the form
\begin{equation}
Q=\d A + d B  \qquad A , B \in \a \label{triv}
\end{equation}
is to be considered as trivial. Notice that in
eq.~(\ref{consist},\ref{triv}) the objects $A,\; B,\; Q,\; R$ are
globally defined on $M$ as differential forms.
 This is the cohomology  of $\d$ modulo
$d$ on $\a$, denoted $H(\d \vert d)$.

The very existence of non-trivial solutions originates in the locality
condition on $Q$.
Indeed relaxing this condition may wipe out the cohomology~\cite{WZ71}.

We first introduce the necessary objects  (certain algebras of polynomials
in the fields, equipped with their differentials) and recall what are
the so-called descent equations. We then calculate $H(\d)$
and the cohomology of $d$ on $H(\d)$.
This step is crucial since it allows to show why the Ansatz made
in~\cite{dvtv85b},
namely that the solutions are built out of the {\it differential forms} $A$ and
$F$ (gauge potential and
curvature) rather than their individual components, does not restrict
the generality of  the  solution, up to elements of $H(\d)$.
We conclude by constructing the {\em independent} elements of $H( \d
\vert d)$, i.e. giving the
general solution of the Wess-Zumino  consistency equation.

We shall assume that the principal fiber bundle where the gauge
potential lives is trivial, i.e. identified with $M\times G$. We thus
avoid the intricacies of handling a reference connection which are
believed to be inessential~\cite{St76}, but we will return to this problem
elsewhere.

Here $M$ is any $n$--dimensional
space--time and $G$ is a compact Lie group. The Lie algebra of $G$ is
denoted $\g$ with structure constants $f^i_{jk}$ in some basis.

\section{Preliminaries}

We denote by $\a$ the algebra of form--valued functionals
$\omega(A,\chi)$ of the fields $A$ (gauge--potential) and $\chi$
(ghost) such that for any $x\in M$, $\omega(A,\chi)(x)$ depends only
on the fields and a finite number of their derivatives at $x$. They
may depend  on external fields such as a metric on $M$.

In a local chart $U$, $\a$ is generated by $\Omega(U)$ the
differential forms on $U$, and the fields $A^i_\mu$, $\chi^i$ and
their derivatives (including in particular the field--strength $F^i_{\mu\nu}$).
We shall use the subalgebra $\b$ of $\a$ generated
by the {\em forms} $A^i,~F^i,~\chi^i,~d\chi^i,~\Omega(M)$, and
the subalgebra $\h$ of $\b$ generated by the $G$--invariant
polynomials $P(\chi)$, $Q(F)$.

The action of the differentials $d$ and $\d$ is easily defined on the
generators of $\a$:
\begin{equation}
d A^i_\mu = A^i_{\mu,\nu}\,dx^\nu ,\quad d\chi^j = \chi^j_{,\nu} dx^\nu
\end{equation}
and so on, where $A^i_{\mu,\nu},~\chi^j_{,\nu},~dx^\nu$ are independent
generators and $d$ is the exterior differential on $\Omega(M)$, while
\begin{equation}
\d A^i_\mu = \chi^i_{,\mu} + f^i_{jk} A^j_\mu\chi^k,\quad
\d \chi^i=-{\textstyle{ 1\over 2}}\; f^i_{jk} \chi^j \chi^k, \quad \d \alpha=0
{}~\forall \alpha \in \Omega(M).
\end{equation}
Then $d,~\d$ are extended as antiderivations of $\a$ in such a way
that~eq.~(\ref{ddel}) is verified.

In order to fix the notations we shall call $\k$ the subalgebra of
$\b$ generated by $\h$ and $\Omega(M)$ and $\k_c$ (resp. $\k_b$) the
subalgebra generated by $\h$ and the closed (resp. $d$--exact) forms on $M$.
They will appear in the computation of $H(\b,\d)$ and $H(H(\d),d)$ in the
following.

One of the main tool used from the early days of the study of the
Wess--Zumino equation are the so called descent equations.
If $Q$ is some representative of
$H(\d |d)$ then eq.~(\ref{consist}) produces for us a local polynomial
$R$ which turns out to be a representative for some element of $H(\d
|d)$. Indeed $R$ verifies $\d(dR)=0$ and thus there exists a local
polynomial $S$ such that $\d R +dS=0$. This is a consequence of the
triviality of $H(d)$ in form degree strictly smaller than $n=\; \mbox{dim}~M$
noticing that $Q$ is of form degree $\leq n$ whence $R$ is of degree
$\leq n-1$.

By definition $\partial$ is defined only in cohomology by:
$\partial[Q]=[R]$. This definition makes sense since if $Q$ is
trivial, i.e. $Q=\d A + dB$ then $\d Q = \d dB$ and $R$ is of the form
$\d B + d C$ that is to say $[R]=0$.
Notice that $\partial$ decreases the form degree by one and
consequently $\partial^{n+1}=0$.

Choose a representative $Q$ for some element of $H(\d |d)$ and let $k$
be the smallest integer such that $\partial^{k+1} [Q]=0$. We may write
the descent equations (or ladder):
\begin{eqnarray}
\d Q + d Q_1 &=&0	\nonumber\\
\d Q_1 + d Q_2 &=&0	\nonumber\\
\dots &&	\label{descent}\\
\d Q_{k-1} + d Q_k &=&0	\nonumber\\
\d Q_k &=&0	\nonumber
\end{eqnarray}

\section{Computation of $H(\d)$}
\label{Hdelta}

Clearly the problem is of a local nature since $\d$ acts trivially
on $\Omega(M)$.
We may thus work in a coordinate patch $U$, where $\a$ is
the tensor product of $\Omega(U)$ with the algebra generated by
$A^i_\mu$, $\chi^i$ and their derivatives. The $\d$ cohomology
becomes obvious if one takes as a system of generators:
\begin{equation}
A^i_{(\mu,\nu_1,\dots,\nu_p)},\quad \d
A^i_{(\mu,\nu_1,\dots,\nu_p)},\quad \chi^i,
\quad \left( D_{(\nu_1} \dots D_{\nu_p} F_{\mu )\nu} \right)^i,
\quad p=0,1,\cdots
\end{equation}
where $_{ (\quad )}$ means symmetrization over indices, and $D_\rho$ means
covariant derivation. This proper choice makes explicit the splitting
of the algebra generated by the fields into a tensor product of
differential algebras. The $\d$--cohomology of the factor generated by
$(A,\d A)$ is trivial. On the factor generated by $(\chi,D^p F)$, $H(\d)$
is the cohomology of $\g$ acting on the module of polynomials in the
components of $F$ and their covariant derivatives.

It is known~\cite{ChEi48,HoSe53} to be isomorphic to the tensor
product of the invariant
forms on $G$ by the $G$--invariant part of the above module.
Finally in the chart $U$, and denoting $\{\cdots\}^G$ the
$G$--invariant part:
\begin{equation}
H(\d)\simeq \{\mbox{polynomials in~}\chi \}^G\otimes
\{\mbox{polynomials in~}F_{\mu \nu},D_\rho F_{\mu\nu},\dots\}^G
\otimes \Omega(U) \label{hdelta}
\end{equation}
This result was  partly stated in~\cite{Ban87}, but this proof inspired by
the one of~\cite{dvtv85} appears to be much simpler.

The algebra appearing in eq.~(\ref{hdelta}) has an intrinsic meaning,
i.e. is invariant under changes of coordinates. Moreover given a
global $\d$--cocycle on $M$, its representatives in this algebra
over two coordinates patches $U_\alpha$ and $U_\beta$ match on
$U_\alpha \cap U_\beta$. Indeed their difference being a
$\d$--coboundary must vanish.
So the global result (on $M$) is immediately obtained by restricting
oneself to objects of the above type globally defined on $M$.
For instance  the lagrangian Tr$(F\wedge \ast F)$ which
explicitely contains the metric on $M$ as an external field belongs
to $H(\d)$.
\proclaim Proposition. $H(\d)$ is the skew--tensor product of the algebra of
invariant polynomials in $\chi$ by the algebra of globally defined
forms on $M$ constructed with invariant polynomials in $F_{\mu\nu}$
and its covariant derivatives.

Similarly the algebra $\b$ is generated by the forms $A^i,\; \d A^i,\;
\chi^i,\;F^i,\; \Omega(M)$ so that $H(\b,\d)\simeq \k$ with $\k$
defined in the previous section. We see that $H(\b,\d)$ is naturally
included in $H(\a,\d)$.

\section{Computation of $H(H(\d),d)$}
\label{computation}

We shall prove here a beautiful result: in form degree smaller than
$n$ {\em only the differential
forms $A,~F$ and  not their individual components nor their
derivatives survive the calculation of the cohomology of $d$ on $H(\d)$.}
As will be shown later, this calculation  validates the older analysis
of~\cite{dvtv85b}.

The computation proceeds in steps:\newline
-- the abelian case ($G=U(1)$) in ghost degree zero\newline
-- abelian case with many $U(1)$ factors and a possible global
symmetry, ghost degree zero.\newline
-- the non abelian case in ghost degree zero.\newline
-- non abelian case, unrestricted ghost degree.\newline
\proclaim Proposition.
$H(H(\d),d)$ is
generated in form degree smaller than
$n$ by $\h$ and $H_{DR}(M)$, where $H_{DR}(M)$ is the de Rham
cohomology of space--time. \par
\noindent {\bf Proof.}

{\em Step 0.}
In ghost degree zero, and for just one abelian potential
$A_\mu$ and its field--strength $F_{\mu\nu}$ we show that
in form degree smaller than $n$,
the cohomology of $d$  on the space of polynomials in $F_{\mu\nu}$ and
its derivatives (with coefficients forms on $M$) is generated by
polynomials in $F=F_{\mu\nu}dx^\mu dx^\nu$ and the de Rham
cohomology of $M$.

In other words suppose $Q$ is a polynomial in the components
of $F$ and their derivatives, and verifies  $d Q=0$. Then
$$Q= d R(F_{\mu\nu},\partial_\rho F_{\mu\nu},\dots) +U$$
where  $U$ is   a polynomial in the form $F$, i.e.
$U = \sum_k (F)^k \; \omega_k$, with the   $\omega_k$  representatives of
the de Rham cohomology of $M$ . The proof uses  the algebra $\a$ and
the descent equation, whose existence comes from the triviality of d
on $\a$.
The last non trivial term $Q_k$ in the ladder is a representative of
$H(\d)$, i.e a polynomial in $\chi$, $F_{\mu\nu}$ and its derivatives
(no  derivatives of $\chi$). Since $\chi^2=0$, $k$ is at most one.

The ladder takes the form:
\begin{eqnarray*}
Q &=& dQ_0(A_\mu, A_{\mu,\nu},\dots) \\
0 &=& \d Q_0 +d Q_1 \\
0 &=& \d Q_1
\end{eqnarray*}
$Q_1$ is necessarily of the form
$Q_1 = \chi \; P(F_{\mu\nu}, \partial_\rho  F_{\mu\nu},\dots )$, from which
$\d Q_0 +d\chi  P - \chi dP =0$. Thus $\d (Q_0 + A
P)=\chi dP$, from which $dP=0$. As a consequence
$$ Q_0 + A P = R(F_{\mu\nu}, \partial_\rho  F_{\mu\nu},\dots ) +
\d S$$
Since $Q_0$ and $P$ are of ghost degree 0, $\d S=0$, and
$Q= d Q_0= dR - F P$.
One gets the desired result by an induction on the degree of $P$ in $F$.
Notice that one gets a polynomial $U$ of the stated form.

During this proof
one does not create non--locality in the fields: if for example $d
Q=0$ and $Q$ depends in a local way on some other auxiliary fields,
$Q$ is $d$ equivalent to some polynomial $Q'$ in $F$ up to quantities of
the form $d R$ with both $Q'$ and $R$ local in these auxiliary fields.

{\em Step 1.}
Let us consider several {\em abelian} fields $A^i_\mu$.\hfil\break
{\em a) Cocycle condition.} Let
$Q(F^i_{\mu\nu},F^i_{\mu\nu,\rho},\dots)$ be a polynomial with
form coefficients such that
$dQ=0$.
To  show that $Q= d R + U(F^i)$ where $R$ is a similar polynomial,
while $U$ involves only the forms $F^i$, we first  write $d=d_1+d'$ with
$d_1$ acting only on $F^1$ and $d'$ acting
on all the other fields. Suppose $Q$ is of maximal order $k$ in $F^1$ (total
number of derivatives of $F^1$). Let $Q= Q^k+ Q'$ where $Q^k$ contains
the terms of order $k$. Since $d_1 Q^k =0$
we may apply the previous result and get
$Q^k=d_1 R_1 + U_1$ where $U_1$ depends on $F^1_{\mu\nu}$ only through
the form $F^1$,  $R_1$ may be chosen
of order at most $(k-1)$ and both are local in $F^2_{\mu\nu},
F^2_{\mu\nu,\rho}, \dots$
In $Q-d R_1$ the field $F_1$ is of order at most $(k-1)$ in $F^1$.
By induction we get  $Q=dR+U(F^1,F^2_{\mu\nu},\dots)$.
The polynomial  $U$ may be expanded in exterior powers of $F^1$:
$U= \sum (F^1)^{^k} \alpha_k $.
Each of the terms $\alpha_k$  is local in $F^2_{\mu\nu},\dots$ and
separately verifies $d\alpha_k=0$ since $dU=\sum (F^1)^{^k} d\alpha_k=
0$ identically in $F^1$, from
which we can proceed similarly for the other fields. Finally:
$$Q=\sum_l Q_l(F) \alpha_l \quad\mbox{with~}\alpha_l\in\Omega(M),
\quad d\alpha_l=0.$$
\hfil\break
{\em b) Coboundary condition.}
We want to solve $U=dV$ with $V$ a polynomial in $F^i_{\mu\nu}$ and
their derivatives with coefficients in $\Omega(M)$ and $U$ some
polynomial in the forms $F^i$. We may split $d$ in  $d=d_F+d'$
where $d_F$ acts only on the fields $F^i_{\mu\nu}$ and $d'$ acts on
the other fields. Let $V^m$ the part of $V$ of maximal order $(m)$ in the
derivatives of $F$ so that $d_F V^m=0$. The previous analysis allows
to write: $V^m = d_F W+Q(F)$
with $Q$ a polynomial in the forms $F^i$. Then $V$ may be replaced by
$V- dW$ which is of order $(m-1)$. One reaches $U=dV$ with $V$ of
order zero. At this point $d_F V=0$ hence $V=\sum_l Q_l(F) \beta_l$ where the
$\beta_l$ are arbitrary forms on $M$, from which:
$$U=dV=\sum_l Q_l(F) d\beta_l.$$
This is the place where the de Rham cohomology of $M$ appears.
Notice that these arguments could be turned into a spectral sequence
argument similar to the one used in~\cite{hdtv91} in the calculation of $H(d)$.
\hfil\break {\em c) Global invariance.}
Assume now that  $G$ acts as a global transformation group of the fields
$F^i$, and suppose $P$ is an invariant polynomial in the $F^i$ such
that $dP=0$.
Then $P=dR + U$. Taking the mean over the group yields $P=d
\overline{R} + \overline{U}$ with $\overline{R}$ and $ \overline{U}$
invariant by $G$. We may do the same for the coboundary condition.

{\em Step 2.}
We may now go to the non-abelian case with local invariance.
Suppose $P$ is a $G$-invariant
polynomial in the $F^i_{\mu \nu} $ and their covariant derivatives,
and that its degree as a form is strictly smaller than $n$.
The cocycle condition
$dP=0$ implies the existence of $G$--invariant $R_1$ and $U_1$ such
that
$$ P = dR_1(F^i_{\mu \nu}, \partial_\rho F^i_{\mu \nu}, \dots ) +
U_1(F^i) $$
Replacing ordinary derivatives by covariant ones
$$  P_1 = P - dR_1(F^i_{\mu \nu}, D_\rho F^i_{\mu \nu}, \dots )$$
is again a $d$--closed invariant polynomial in  the
$F^i_{\mu \nu} $ and their covariant derivatives, but of lower order
of derivatives of $F$. We will arrive at the conclusion  that
$$ P = dR(F^i_{\mu \nu}, D_\rho F^i_{\mu \nu}, \dots ) + U(F)$$
with $R$ and $U$ being  $G$--invariant. The coboundary condition may
be analyzed along the same lines.

To sum up this discussion, we see that cocycles in form degree
strictly smaller than $n$ are cohomologically equivalent to some
$U(F)=\sum_l P_l(F) \alpha_l$. Here $P_l(F)$ are independent
invariant polynomials
of the forms $F^i$ with numerical coefficients and $\alpha_l$
closed differential forms on $M$. Such a cocycle is trivial, i.e.
$U = d V +\d W$ and $V \in H(\d)$
if and only if:
$\alpha_l = d \beta_l$ (of course $\d W$ is irrelevant here since
we are living in ghost degree zero).

{\em Step 3.}
To treat the situation with non zero ghost degree, one first notices
that the action of $d$ on invariant polynomials in $\chi$ vanishes in
$H(\d)$. Indeed for such a $P(\chi)$ there exists some $Q \in \b$ such
that $\d Q + d P=0$. Consequently  $H(\d)$ is a tensor product
of a purely ghost part
(invariant polynomials in $\chi$) on which $d$ vanishes, by the part
which we have just analyzed. By K\" unneth theorem, the cohomology
of $d$ on $H(\d)$ is thus
obtained from the one we have just calculated by taking its tensor
product with the algebra of invariant polynomials in $\chi$ (i.e.
invariant forms on $\g^*$). In dimension strictly smaller than $n$ the
cycles (resp. boundaries) of the action of $d$ on $H(\d)$ may be
identified with elements of $\k_c$ (resp. $\k_b$) keeping in mind that
$d$ is the induced $d$ on $H(\d)$.

It is straightforward to see that the computation of $H(H(\b,\d ),d)$
leads to the same answer, i.e. the cycles are identified with $\k_c$
and the boundaries to $\k_b$ in dimension smaller than $n$.

\section{Computation of $H(\d | d)$}

\subsection{Cocycle condition.}
We want to solve eq.~(\ref{consist}), by analyzing in some detail the
descent homomorphism $\partial$ of the ladder~(\ref{descent}).
What we want to show is  that we may choose representatives
$Q_k,\; Q_{k-1},\dots,Q_1$ in the algebra $\b$ rather than $\a$.
The idea is to go up from the bottom of the ladder and fix the choice
of these representatives. This will
permit us to make contact with our results of~\cite{dvtv85b,dvtv86}
and produce the
desired cohomology.

If $k=0$, $Q$ is in $H(\d)$ and the general solution of the condition
$\d Q=0$ is known (see section~\ref{Hdelta}).

If $k \geq 1$ $Q_k$ verifies:
\begin{eqnarray}
\d Q_{k-1} + d Q_k &=&0  \label{qk-1}   \\
\d Q_k &=&0  \label{qk}
\end{eqnarray}
The previous equations indicate that $Q_k$ is a $d$--cocycle in $H(\d)$.
Since by hypothesis on $k$, $Q_k$ is non trivial in $H(\d |d)$ it
cannot be trivial in $H(H(\d),d)$. It is equivalent in $H(H(\d),d)$ to
some element $B_k$ of $\k_c$. We may choose $Q_k=B_k$.

We have $\d(\d Q_{k-1} + dB_k)=\d (dB_k)=0$ hence there exists
$B_{k-1}\in \b$ and $X$ a representative of $H(\b,\d)$ in $\k$ such that
\begin{equation}
\d B_{k-1} + d B_k =X \label{x}
\end{equation}
{}From equations~(\ref{qk-1},\ref{x}) we see that
\begin{equation}
\d(B_{k-1}-Q_{k-1})=X \label{xx}
\end{equation}
meaning that $X$ is trivial in $H(\a,\d)$. Since we have a natural
{\em inclusion} of $H(\b,\d)$ in $H(\a,\d)$ this forces $X$ to vanish.
In conclusion there exists $B_{k-1}$ in $\b$ such that:
\begin{equation}
\d B_{k-1} + d B_k =0. \label{x0}
\end{equation}

If $k=1$, then $\d(Q-B_0)=0$ and $Q$ is the sum of an arbitrary
$\d$--cocycle and of some element $B_0$ in $\b$ satisfying the
consistency condition. These elements have been completely described
in~\cite{dvtv85b}.

If $k \geq 2$ we are going to show that $Q_{k-1}$ may be taken in
$\b$. Indeed there exists $Q_{k-2}$ such that
\begin{equation}
\d Q_{k-2} + d Q_{k-1} =0  \label{qk-2}
\end{equation}
Since from eq.(\ref{x0}), $\d(dB_{k-1})=0$, there exists $B_{k-2}\in\b$
and $Y$ a representative of $H(\b,\d)$ in $\k$ such that
\begin{equation}
\d B_{k-2}+d B_{k-1}=Y. \label{y}
\end{equation}
Equations (\ref{xx},\ref{qk-2},\ref{y}) yield the conditions:
\begin{eqnarray}
\d (B_{k-2}-Q_{k-2})+d(B_{k-1}-Q_{k-1})&=&Y \label{bqy}\\
\d(B_{k-1}-Q_{k-1})&=&0\label{bq}
\end{eqnarray}
The previous equations indicate that $Y$ is trivial in $H(H(\d),d)$
and is thus of the form $Y=\sum_l d\beta_l \wedge P_l$ with $\beta_l
\in \Omega(M),\quad P_l \in \h$.
It is possible to construct $R_l \in \b$ such that $dP_l=\d R_l$, from
what we know of $H( H(\b,\d),d)$.

The replacement of $B_{k-1}$ by $B_{k-1}-\sum_l \beta_l
\wedge P_l$  and of $B_{k-2}$ by $B_{k-2} - \sum_l \beta_l
\wedge R_l$ leaves eq.~(\ref{x0},\ref{bq}) unchanged
while eq.~(\ref{bqy}) becomes
\begin{equation}
\d (B_{k-2}-Q_{k-2})+d(B_{k-1}-Q_{k-1}) =0.  \label{bq0}
\end{equation}
{}From equations~(\ref{bq0},\ref{bq}) we see that  $(B_{k-1}-Q_{k-1})$
is of the form $U+\d R +dS$ with $U$ in $\k_c$ and $\d S=0$. If we
redefine  again
$B_{k-1}$ as $B_{k-1} - U$ we see that the class $[Q_{k-1}]$ in
$H(\d | d)$ has a representative in $\b$ (namely $B_{k-1}$).
\hfil\break
{\bf Remark.} We see here how going up the descent equation is not
uniquely defined: there is an ambiguity due to the non triviality of
$H(\d)$. This ambiguity appears in the freedom of choice of $B_{k-1}$.
The successive redefinitions of $B_{k-1}$ may change its class in
$H(\d | d)$.

Now we can set $Q_{k-1}=B_{k-1}$ and change $Q_{k-2}$ into
$Q_{k-2}+dR$ and no further modification of the ladder. We are brought
back to the same situation with $k$ replaced by $(k-1)$. Without
further ado we see that $Q$ is equivalent to the sum of an
arbitrary $\d$--cocycle
and one of the known solutions in $\b$ of the consistency condition.

Notice that, in the course of the proof, we have proved (and used)
 the fact that if some $Y\in
\k$ verifies $Y= \d A + d H$ with $A \in \a$ and $H \in H(\d)$, then
there exist $\alpha$ in $\b$  and $\beta$ in $\k$,
such that $Y= \d \alpha + d \beta$.

\subsection{Coboundary condition.}

Since the solutions of the consistency condition are defined up to a
$\d$--cocycle we start from a solution belonging to $\b$. In
reference~\cite{dvtv85b,dvtv86} we have produced the
list of all cohomologically {\em independent}
solutions of the cocycle condition in $\b$.

We shall now see that a
non trivial solution in $\b$ remains non trivial in $\a$. Let us take
$Q \in \b$ and the smallest integer $k$ such that $\partial^{k+1} Q
=0$.
 By hypothesis $k \geq 0$ and $P = \partial^{k} Q \in \k$ is non trivial
for the cohomology of $\d$ modulo $d$ computed in $\b$. We shall prove
that $P$ remains non trivial in $A$ hence $[Q]$ is not zero in $\a$
since $[P]=\partial^k [Q]$ in $\a$.
\proclaim Proposition.
If $P$ in $\k$ is  of the form $P=\d A +d B$, with $ \quad A,B \in\a$, then
there exist $\alpha$ and $\beta$ in $\b$ such that
$P=\d\alpha+d\beta$.\par
\noindent {\bf Proof.}
Since $\d P=0$, we know that  $d(\d B)=0$ and thus
there exists $C \in \a$ such that $\d B+dC=0$, i.e. $B$
verifies the consistency
condition we have just solved. We know $B$ is of the form:
$B=G + \d F + H$ where $G \in \b$ is a solution of the
consistency condition, $F \in \a$ and $H$ is a representative of
$H(\d)$.
{}From this
$$P-d G = \d (A - d F) + d H$$
 showing that $\d( P-dG)=0$.
As a consequence, $(P-dG)$ is  a $\d$--cocycle of $\b$, and may thus be
written $P-dG=X+\d Y$, with $X\in \k$, and $Y\in \b$.

Setting $A'=A - d F - Y$,we get $\d A' + d H=X$. We know from the last
remark of the previous section that this implies
the existence of $U$ and $V$ in $\b$ such that
$X=\d U + d V$, concluding the proof   if we set $\alpha= Y +U  $ and
$ \beta = G +V$.

\section{Conclusion}

We have produced the calculation of various related cohomologies.
The $\d$--cohomology has been easily calculated, and we have shown the
r\^ ole played by the cohomology of $d$ on $H(\d)$. We have in
particular shown where the de Rham cohomology of space--time enters
the calculations exactly as it is the case in the evaluation of the de
Rham cohomology of the orbit space of connections~\cite{AtBo82,AsMi86}.

By bringing back the problem to the
one solved in~\cite{dvtv85b}, we have shown that up to the addition of a
non $d$--trivial
$\d$--cocycle in $\a$, the general solutions are obtained applying
a ``generalized transgression'' to products of elements of $\h$ with
representatives of the de Rham cohomology of space--time.

\end{document}